\documentclass[12pt]{article}
\usepackage{amsfonts}
\usepackage{amsmath}
\usepackage{amssymb}
\usepackage{amscd}
\usepackage[dvips]{graphicx}

\def\HH{\mathbb{H}}
\begin{document}

\begin{center}
\bigskip

\bigskip

$\mathbf{\kappa -}${\Large \textbf{Deformed Covariant Quantum Phase Spaces
as Hopf Algebroids}}
\end{center}

\bigskip

\begin{center}
{{\large ${\mathrm{Jerzy\;Lukierski}}^a$, ${\mathrm{Zoran\;\check{S}koda}}{}^b$,
${\mathrm{Mariusz\;Woronowicz}}^a$ }}

\bigskip

{
${}^a\mathrm{Institute\;for\;Theoretical\;Physics}$}

{
$\mathrm{\ University\; of\; Wroclaw\; pl.\; Maxa\; Borna\; 9,\; 50-204\;
Wroclaw,\; Poland}$}

\bigskip

{
${}^b\mathrm{\ Faculty\; of\; Science,\; University\; of\; Hradec\; Kr\acute{a}lov\acute{e}}$

$\mathrm{\ Rokitansk\grave{e}ho \; 62,\; Hradec\; Kr\acute{a}lov\acute{e},\; Czech\; Republic}$}

\end{center}
\begin{abstract}
We consider the general $D=4$ $(10+10)$-dimensional $\kappa $-deformed
quantum phase space as given by Heisenberg double $\mathcal{H}$\ of $D=4$ $%
\kappa $-deformed Poincar\'e-Hopf algebra $\HH$. The standard $(4+4)$%
-dimensional $\kappa$-deformed covariant quantum phase space spanned by $%
\kappa $-deformed Minkowski coordinates and commuting momenta generators $(%
\widehat{x}_{\mu },\widehat{p}_{\mu })$ is obtained as the subalgebra of $%
\mathcal{H}$. We study further the property that Heisenberg double defines\
particular quantum spaces with Hopf algebroid structure. We calculate by
using purely algebraic methods the explicit Hopf algebroid structure of
standard $\kappa $-deformed quantum covariant phase space in Majid-Ruegg
bicrossproduct basis. The coproducts for Hopf algebroids are not unique,
determined modulo the coproduct gauge freedom. Finally we consider the
interpretation of the algebraic description of quantum phase spaces as Hopf
algebroids.
\end{abstract}
\section{Introduction}

Recently several papers appeared (see e.g \cite{1}-\cite{5}) discussing the
bialgebroid and Hopf algebroid structures of deformed quantum phase spaces
with noncommutative coordinates satisfying the $\kappa $-deformed Minkowski
space-time algebra \cite{6}-\cite{8}. In these considerations the covariance
under the action of $\kappa $-deformed quantum Poincar\'e symmetries was not
properly exposed\footnote{%
Following \cite{7} the covariance of $\kappa $-Minkowski algebraic relations
under the action of $\kappa $-deformed quantum Poincar\'e algebra is an
unseparable part of the definition of $\kappa $-deformed noncommutative
Minkowski space.}. However, to obtain $\kappa $-deformed quantum phase space
with built-in quantum $\kappa $-covariance property it is convenient to
employ the Heisenberg double construction, which in a given $\kappa $%
-Poincar\'e algebra basis leads to unique choice of covariant $\kappa $%
-deformed quantum phase space algebra. Further using the property that
Heisenberg double algebra defines a Hopf algebroid, we shall point out here
some new properties of the quantum phase spaces equipped with coalgebraic
sector.

The notion of Hopf algebroid introduces new class of quantum spaces which
provide for deformed quantum phase spaces the bialgebraic structure \cite{9}-%
\cite{12} with a freedom in coproducts which will be called the coproduct
gauge. The mathematical origin of such a freedom is linked with the
bialgebroid structure of\ $\mathcal{H}$ described briefly as follows:

\begin{enumerate}
\item Algebraic sector of $\mathcal{H}$ is given by the total algebra $H$,
and its subalgebra $A$ called the base algebra,

\item there exist two maps: the source map $\alpha :A\longrightarrow H$
which is algebra homomorphism and
the target map $\beta :A\longrightarrow H$
which is algebra antihomomorphism.
The images of the two maps from $A$ into $H$\ commute in $H$,
i.e.\ for any $a,b\in A$
\begin{equation}
\lbrack \alpha (a)\,,\beta (b)]=0,  \label{1}
\end{equation}

what permits an $(A,A)$-bimodule structure on $H$, namely
$a.h.b = h\beta(a)\alpha(b)$.

\item For bialgebroids the notion of Hopf-algebraic coproducts $\Delta
:H\longrightarrow H\otimes H$\ \ ($\otimes $ describes standard tensor
product) is replaced by the coproduct map from $H$ into $(A,A)$-bimodule
product $H\otimes _{A}H$, with respect to the above bimodule structure on
$H$ (see \cite{10},\cite{12}).
It appears that $H\otimes_A H$ as the codomain of
the co-algebraic sector of bialgebroids
does not inherit the algebra structure from $H\otimes H$\ \footnote{%
The factorwise algebra multiplication is however well defined in
the Takeuchi product \cite{5},\cite{12}, certain subbimodule
$H\times_A H\subset H\otimes_A H$ introduced
by Takeuchi \cite{9}. The coproduct $\Delta$
is required to take values within $H\times_A H$ and
$\Delta:H\to H\times_A H$ must respect the multiplication.
}. $H\otimes _{A}H$ can be
defined as the quotient of $H\otimes H$ by the left ideal $\mathcal{I}_{L}$\
generated as a left ideal
by the subset of $H\otimes H$ consisting
of all elements of the form $\alpha(a)\otimes 1 -
1\otimes \beta(a)$ \cite{10}\footnote{%
In fact the bialgebroid structure obtained with the use of left ideal (2)
defines a right bialgebroid $\mathcal{H}^{R}$ (see e.g. \cite{11}-\cite{13},%
\cite{5}). The choice of the right bialgebroid in this paper is related with
further use of coproduct formulae given in \cite{14}.},
\begin{equation}
\mathcal{I}_{L}=\langle
\alpha (a)\otimes 1-1\otimes \beta (a),\,\,\,\,\, a\in A\rangle.
\label{2a}
\end{equation}
\end{enumerate}

If we introduce the canonical choice $\alpha (a)=a$, one gets the left ideal
in the form
\begin{equation}
\mathcal{I}_{L}=\langle a\otimes 1-1\otimes \beta (a),\,\,\,\,\,a\in A\rangle,
\label{3}
\end{equation}%
with the target map $\beta$ (with $\beta (a)\in H$) determining the coalgebra gauge freedom.

If the base algebra is commutative (e.g.\ for canonical Heisenberg
algebra, see \cite{1}-\cite{4}) and for some other special classes of Hopf
algebroids, one can introduce the coproduct gauge as
defined by two-sided ideal, namely the bialgebroid coproduct
takes values in the standard tensor product $H\otimes H$
divided by a two-sided ideal $\mathcal{I}\subset H\otimes H$.

We recall that in Hopf-algebraic $\kappa $-deformation scheme the general
covariant $\kappa $-deformed phase space is provided by the Heisenberg double%
\footnote{%
Heisenberg double $\mathcal{H}$ is a special case of smash product algebra
$\HH\rtimes V$, where $V$ is an $\HH$-module algebra,
which is in general case not
endowed with the Hopf-algebraic structure (see e.g.~\cite{25}).
\par
{}} $\mathcal{H}=\HH\rtimes \widetilde{\HH}$ (see e.g.~\cite{16}), where $%
\HH=U_{\kappa }(\widehat{g})$ describe $\kappa $-deformed Poincar\'e-Hopf
algebra \cite{13a},\cite{7}\ and $\widetilde{\HH}$ is the Hopf algebra
describing dual $\kappa $-deformed quantum Poincar\'e group. In this paper we
employ the general property (see e.g.~\cite{10}, Sect. 6) that Heisenberg
double algebra is equipped with the Hopf algebroid structure. In recent
literature (see e.g.~\cite{2},\cite{3}) the bialgebroid structures of
deformed standard quantum phase spaces $(\widehat{x}_{\mu },\widehat{p}_{\mu
})$ with $\kappa $-Minkowski space-time sector%
\begin{equation}
\lbrack \widehat{x}_{0},\widehat{x}_{i}]=-\frac{i}{\kappa }\widehat{x}%
_{i},\qquad \lbrack \widehat{x}_{i},\widehat{x}_{j}]=0,  \label{4}
\end{equation}%
and commuting fourmomenta $\widehat{p}_{\mu }$ were studied by embedding
into canonical quantum phase space algebra (we put $\hslash =1$)%
\begin{equation}
\lbrack x_{\mu },x_{\nu }]=[p_{\mu },p_{\nu }]=0,\qquad \lbrack x_{\mu
},p_{\nu }]=i\eta _{\mu \nu }.  \label{5}
\end{equation}%
Relation (\ref{4}) permits the following general class of realizations of
quantum phase spaces\footnote{%
Such realizations can be expressed by differential operators, with
$p_\mu$ replaced by ($\sqrt{-1}$ times) the partial derivatives $\partial_\mu$
\cite{13d},\cite{1}-\cite{5}.
{}}%
\begin{equation}
\widehat{x}_{\mu }=x_{\nu }f_{\,\mu }^{\nu }(p),\qquad \widehat{p}_{\mu
}=p_{\mu },  \label{6}
\end{equation}%
where $f_{\,\mu }^{\nu }(p)$ are chosen in consistency with relations (\ref%
{4}),(\ref{5}) (Jacobi identities) and provide large variety of quantum
phase spaces with space-time algebra described by relations (\ref{4}). The
problem with such an approach is the lack of structural indication how to\
obtain the covariant action of $\kappa $-deformed Poincar\'e-Hopf algebra,
which is a part of full definition of quantum $\kappa $-deformed Minkowski
space (see $^{1}$).

In this paper we use the construction of the $\kappa $-deformed quantum
phase space as the Heisenberg double of $D=4$ $\kappa $-deformed Poincar\'e-Hopf algebra first presented in \cite{14}. Such construction contains built-in $%
\kappa $-covariance of $\kappa $-deformed quantum phase space first
observed for $\kappa $-Minkowski space-time sector in \cite{7}. In
Majid-Ruegg basis \cite{7}\ we obtain that both $\kappa $-Poincar\'e-Hopf
algebra $\HH$ and $\kappa $-Poincar\'e group $\widetilde{\HH}$ are described by
two dual bicrossproduct structures \cite{14},\cite{16},\cite{23},\cite{13c},
namely\footnote{
It follows from formula (\ref{6a}) that the $\kappa $-deformation in algebraic
sector of $\HH$ is present only in cross commutators between fourmomenta and
Lorentz generators. From (\ref{6a}), it follows that $\HH$ can be described by the
action of $U(o(1,3))$ on $\mathcal{T}^{4}$ as well as the coaction of $%
\mathcal{T}^{4}$ on $U(o(1,3))$.}
\begin{equation}
\HH=U(so(1,3)){\triangleright \!\!\!\blacktriangleleft }_{\kappa }\mathcal{T}%
^{4}\qquad \overset{duality}{\longleftrightarrow }\qquad \widetilde{\HH}=%
\widetilde{\mathcal{T}}_{\kappa }^{4}{\triangleright
\!\!\!\blacktriangleleft }_{\kappa }\mathcal{L}^{6},  \label{6a}
\end{equation}%
where $\mathcal{L}^{6}$ describe the functions of Abelian Lorentz parameters
$\lambda_{\mu \nu }^{\quad }$ which are dual to $U(so(3,1))$ and $\mathcal{T}%
^{4}$ is the fourmomenta sector dual to the algebra $\widetilde{\mathcal{T}}%
_{\kappa }^{4}$ describing noncommutative functions of $\kappa $-deformed
Minkowski coordinates (see (\ref{4})). One can show that the $\kappa $%
-Poincar\'e covariance of fourmomentum sector $\mathcal{T}^{4}$
can be derived from the bicrossproduct structure of $\HH$ (see (\ref{6a})).

The $\kappa $-deformed Poincar\'e algebra $\HH$ acts on standard $\kappa $%
-deformed quantum phase space $\widetilde{\mathcal{T}}_{\kappa }^{4}\otimes
\mathcal{T}^{4}$ in a covariant way. Further, the covariant action of $\HH$ on
$\mathcal{L}^{6}$ follows from the duality of $\mathcal{L}^{6}$ and $%
U(so(3,1))$ algebras as well as the semidirect product of the coalgebra
sectors in $\HH$ and $\widetilde{\HH}$.

Firstly, in Sect.~2, we recall the results presented in \cite{14} and
provide the $10+10$-dimensional generalized $\kappa $-deformed quantum phase
space, with standard dual pair of generators $(\widehat{x}_{\mu },\widehat{p}%
_{\mu })$ and the dual canonical pair $(\widehat{\lambda }_{\mu \nu }^{\quad
},\widehat{m}_{\mu \nu })$ of Lorentz group parameters and Lorentz algebra
generators. In such a way we obtain the $\kappa $-deformation of canonical
generalized phase space which in undeformed case was used for the geometric
description of elementary particles with translational and spin degrees of
freedom (see e.g.~\cite{17}-\cite{21}).

In present paper, in order to provide the explicit formulae describing Hopf
bialgebroid structure, we shall restrict our considerations to standard $%
\kappa $-deformed quantum phase space for spinless particles, given by
Heisenberg double $\mathcal{H}^{(4,4)}\equiv \mathcal{H}_{(p,x)}=\HH_{p}%
\rtimes \HH_{x}$, where the dual Hopf algebras $\HH_{p},\HH_{x}$ describe momenta
and coordinate sectors%
\begin{eqnarray}
\HH_{p} &:&\qquad \lbrack \widehat{p}_{\mu },\widehat{p}_{\nu }]=0,\qquad
\Delta (\widehat{p}_{i})=\ \widehat{p}_{i}\otimes e^{-{\frac{\widehat{p}_{0}%
}{\kappa }}}\ +\ 1\otimes \widehat{p}_{i},\qquad \Delta (\widehat{p}_{0})=%
\widehat{p}_{0}\otimes 1+1\otimes \widehat{p}_{0},  \label{7} \\
\HH_{x} &:&\qquad \lbrack \widehat{x}_{0},\widehat{x}_{i}]=-\frac{i}{\kappa }%
\widehat{x}_{i},\qquad \lbrack \widehat{x}_{i},\widehat{x}_{j}]=0,\qquad
\Delta (\widehat{x}_{\mu })=\widehat{x}_{\mu }\otimes 1+1\otimes \widehat{x}%
_{\mu }.  \label{7a}
\end{eqnarray}

A Hopf algebroid is defined as a bialgebroid with an antipode. In Sect.~3, we
derive the Hopf algebroid structure of standard $\kappa $-deformed quantum
phase space $\mathcal{H}_{(p,x)}$. We determine from formula (\ref{1})
target map (we choose $\alpha (a)=a$) and antipodes; further calculate the
coalgebra gauge sector using two alternative ways of determining tensor
product $H\otimes_{A}H$ over noncommutative ring $A$.
In Sect.~4, we present the
interpretation of particular coproduct gauges in the case of nonrelativistic
QM phase space and we discuss briefly the dependence of results on the
choice of $\kappa $-Poincar\'e algebra basis.

\section{Covariant $\protect\kappa $-deformed quantum phase spaces as
Heisenberg doubles}

\subsection{Heisenberg double - general remarks}

The name "Heisenberg" originates from the simple example of Heisenberg
algebra in quantum mechanics, which is the Heisenberg double for dual pair
of Abelian Hopf algebras, describing respectively commuting
quantum-mechanical momenta and coordinates (see (\ref{7}), (\ref{7a}) in the
limit $\kappa \longrightarrow \infty $). Heisenberg double construction in
more general case represents algebraic generalization of the notion of
quantum cotangent double for algebraic quantum groups, and provides new
models of deformed quantum phase spaces in QM.

A Hopf algebra $\HH=(A,m,\Delta ,S,\epsilon )$ is a bialgebra (with
multiplication $m:A\otimes A\longrightarrow A$, comultiplication $\Delta
:A\longrightarrow A\otimes A$ and counit $\epsilon$) supplemented
with antipode (coinverse) $S$.
Hopf algebra duality between $\HH$ and $\widetilde{\HH}=(A^{\ast
},m^{\ast },\Delta ^{\ast },S^{\ast },\epsilon ^{\ast })$ requires the
existence of bilinear pairing
$A\otimes A^{\ast }\longrightarrow\mathbb{C}$ denoted $%
<a,a^{\ast }>$ $(a\in A,a^{\ast }\in A^{\ast })$, realizing
vector space duality $A\longleftrightarrow A^{\ast }$
and relating the multiplication
(comultiplication) in $\HH$ with comultiplication (multiplication) in $%
\widetilde{\HH}$ \cite{16},\cite{13d}. For dual Hopf algebras one can introduce the natural action $%
\HH\vartriangleright \widetilde{\HH}$%
\begin{equation}
a\vartriangleright a^{\ast }=a_{(1)}^{\ast }<a,a_{(2)}^{\ast }>,  \label{9}
\end{equation}%
where we use the coproduct notation $\Delta (x)=x_{(1)}\otimes x_{(2)}$ $%
(x=a,a^{\ast })$. The action (\ref{9}) if applied to the product of $a^{\ast
},b^{\ast }\in \widetilde{\HH}$ satisfies the Hopf-algebraic consistency
condition \cite{16}%
\begin{equation}
a\vartriangleright (a^{\ast }b^{\ast })=a_{(1)}^{\ast }<a_{(1)},a_{(2)}^{\ast
}> b_{(1)}^\ast<a_{(2)}, b_{(2)}^\ast> \, = (a_{(1)}\vartriangleright a^\ast)(a_{(2)}\vartriangleright b^\ast),  \label{10}
\end{equation}%
i.e.\ the algebra $A^{\ast }$ is an $\HH$-module algebra. From relations (\ref{9})-(\ref%
{10}) one can deduce the cross multiplication rule in the algebra $\mathcal{A%
}=A\oplus A^{\ast }$%
\begin{equation}
(a\otimes 1)(1\otimes a^{\ast })=a_{(1)}^{\ast }<a_{(1)},a_{(2)}^{\ast
}>a_{(2)},  \label{11}
\end{equation}%
which completes the multiplication rule in Heisenberg double algebra $%
\mathcal{A}$. The algebra $\mathcal{A}$ with cross multiplication rule (\ref%
{11}) defines the Heisenberg double $\mathcal{H}=\HH\rtimes\widetilde{\HH}$ with
Hopf algebroid structure and provides in noncommutative geometry a class of
important examples of quantum spaces with supplemented coalgebra sector.

Following \cite{14}, we exhibit below the Heisenberg double $\mathcal{H}%
^{(10,10)}$ of $\kappa $-deformed Poincar\'e group and its dual $\kappa $%
-deformed \ Poincar\'e algebra\footnote{%
We consider the $\kappa $-deformed Poincar\'e-Hopf algebra in bicrossproduct
basis (see e.g.~\cite{7}), with classical Lorentz algebra. For some remarks
about the dependence of results on the choice of quantum algebra basis see
Sect.~4.}, in order to get the general
$\kappa $-deformed quantum phase
space containing both the translational and Lorentz sectors. The standard $%
\kappa $-deformed quantum phase space with generators $(\widehat{x}_{\mu },%
\widehat{p}_{\mu })$
is a subalgebra $\mathcal{H}^{(4,4)}\subset \mathcal{H}^{(10,10)}$.

\subsection{General covariant $\protect\kappa $-deformed quantum phase space}

\subsubsection{$\protect\kappa $-Poincar\'{e}-Hopf algebra $\HH$}

The $\kappa $-Poincar\'{e}-Hopf algebra $\HH$ in bicrossproduct basis \cite{7},%
\cite{8} has the following form (with conventions
$\mu ,\nu ,\lambda ,\sigma
=0,1,2,3;\,i,j=1,2,3$ and $g_{\mu \nu }=(-1,1,1,1))$\medskip \footnote{%
We denote the $\kappa $-Poincar\'e algebra generators by $(\widehat{p}%
_{\mu },\widehat{m}_{\mu \nu })$ and set $\hslash =1$.}\newline
-\textit{algebra sector}:\medskip
\begin{eqnarray}
\lbrack \widehat{m}_{\mu \nu },\widehat{m}_{\lambda \sigma }] &=&i\left(
g_{\mu \sigma }\widehat{m}_{\nu \lambda }+g_{\nu \lambda }\widehat{m}_{\mu
\sigma }-g_{\mu \lambda }\widehat{m}_{\nu \sigma }-g_{\nu \sigma }\widehat{m}%
_{\mu \lambda }\right)  \notag \\
\lbrack \widehat{m}_{ij},\widehat{p}_{\mu }] &=&-i\left( g_{i\mu }\widehat{p}%
_{j}-g_{j\mu }\widehat{p}_{i}\right)  \label{kapp} \\
\lbrack \widehat{m}_{i0},\widehat{p}_{0}] &=&i\widehat{p}_{i}\medskip
\newline
,\qquad \lbrack \widehat{p}_{\mu },\widehat{p}_{\nu }]=0\medskip \newline
\notag \\
\lbrack \widehat{m}_{i0},\widehat{p}_{j}] &=&i\delta _{ij}\left( \kappa
\sinh ({\frac{\widehat{p}_{0}}{\kappa }})e^{-{\frac{\widehat{p}_{0}}{\kappa }%
}}+{\frac{1}{2\kappa }}\overrightarrow{\widehat{p}}^{2}\right) -{\frac{i}{%
\kappa }}\widehat{p}_{i}\widehat{p}_{j}  \notag
\end{eqnarray}%
$\hfill $\medskip \newline
-\textit{coalgebra sector}:\medskip
\begin{eqnarray}
\Delta (\widehat{m}_{ij})\ &=&\ \widehat{m}_{ij}\otimes I\ +\ I\otimes
\widehat{m}_{ij}\medskip \newline
\notag \\
\Delta (\widehat{m}_{k0})\ &=&\ \widehat{m}_{k0}\otimes e^{-{\frac{\widehat{p%
}_{0}}{\kappa }}}\ +\ I\otimes \widehat{m}_{k0}\ +\ {\frac{1}{\kappa }}%
\widehat{m}_{kl}\otimes \widehat{p}_{l}\medskip \newline
\label{kapp2} \\
\Delta (\widehat{p}_{0})\ &=&\ \widehat{p}_{0}\otimes I\ +\ I\otimes
\widehat{p}_{0}\medskip \newline
\notag \\
\Delta (\widehat{p}_{k})\ &=&\ \widehat{p}_{k}\otimes e^{-{\frac{\widehat{p}%
_{0}}{\kappa }}}\ +\ I\otimes \widehat{p}_{k}  \notag
\end{eqnarray}%
-\textit{counits and antipodes}:\medskip
\begin{eqnarray}
S(\widehat{m}_{ij})\ &=&\ -\widehat{m}_{ij}\medskip \newline
,\qquad S(\widehat{m}_{i0})\ =\ -\widehat{m}_{i0}\ +\ \frac{3i}{2\kappa }%
\widehat{p}_{i}  \notag \\
S(\widehat{p}_{i})\ &=&\ -e^{{\frac{\widehat{p}_{0}}{\kappa }}}\widehat{p}%
_{i},\qquad S(\widehat{p}_{0})\ =\ -\widehat{p}_{0}\medskip \newline
\label{kapp3} \\
\epsilon (\widehat{p}_{\mu }) &=&\epsilon (\widehat{m}_{\mu \nu })=0.  \notag
\end{eqnarray}%
\hfill \medskip

\subsubsection{$\protect\kappa $-Poincar\'{e} quantum group $\widetilde{\HH}$}

Using the following canonical form of duality relations%
\begin{equation}
<\widehat{x}^{\mu },\widehat{p}_{\nu }>\ =\ i\delta _{\nu }^{\mu }\quad <%
\widehat{\lambda }{^{\mu }}_{\nu },\widehat{m}_{\lambda \rho }>\ =\ i(\delta
_{\lambda }^{\mu }g_{\nu \rho }\ -\ \delta _{\rho }^{\mu }g_{\nu \lambda })
\label{dual}
\end{equation}%
we obtain the commutation relations defining $\kappa $-Poincar\'{e} group
\cite{6,23} in the following form\medskip \newline
-\textit{algebra sector}:\medskip
\begin{eqnarray}
\lbrack \widehat{x}^{\mu },\widehat{x}^{\nu }]\ &=&\ {\frac{i}{\kappa }}%
(\delta _{0}^{\mu }\widehat{x}^{\nu }-\delta _{0}^{\nu }\widehat{x}^{\mu
})\medskip \newline
,\qquad \lbrack \widehat{\lambda }_{\nu }^{\mu },\widehat{\lambda }_{\beta
}^{\alpha }]\ =\ 0  \notag \\
\lbrack \widehat{\lambda }_{\nu }^{\mu },\widehat{x}^{\lambda }]\ &=&\ -{%
\frac{i}{\kappa }}\left( (\widehat{\lambda }_{0}^{\mu }-\delta _{0}^{\mu })%
\widehat{\lambda }_{\nu }^{\lambda }-(\widehat{\lambda }_{\nu }^{0}-\delta
_{\nu }^{0})g^{\mu \lambda }\right) \medskip \newline
\label{grup}
\end{eqnarray}%
$\hfill $\newline
-\textit{coalgebra sector}:\medskip
\begin{eqnarray}
\Delta (\widehat{x}^{\mu })\ &=&\ \widehat{\lambda }{^{\mu }}_{\rho }\otimes
\widehat{x}^{\rho }\ +\ \widehat{x}^{\mu }\otimes I\medskip \newline
\label{kop} \\
\Delta (\widehat{\lambda }{^{\mu }}_{\nu })\ &=&\ \widehat{\lambda }{^{\mu }}%
_{\rho }\otimes \widehat{\lambda }{^{\rho }}_{\nu }  \notag
\end{eqnarray}%
-\textit{antipodes and counits}:\medskip
\begin{eqnarray}
S(\widehat{\lambda }{^{\mu }}_{\nu })\ &=&\ \widehat{\lambda }_{\nu }{^{\mu
}\qquad }S(\widehat{x}^{\mu })\ =\ -\ \widehat{\lambda }_{\nu }{^{\mu }}%
\medskip \newline
\widehat{x}^{\nu }  \label{ant} \\
\epsilon (\widehat{x}^{\mu }) &=&0\qquad \ \ \ \ \epsilon (\widehat{\lambda }%
{^{\mu }}_{\nu })=\delta {^{\mu }}_{\nu }  \notag
\end{eqnarray}%
$\hfill $

In the Heisenberg double algebra
$\mathcal{H}^{(10,10)}=\HH\rtimes\widetilde{\HH}$ the commutation
relations (\ref{kapp}) and (\ref{grup}) are supplemented by the following
relations obtained from (\ref{dual}),(\ref{kapp2}) and (\ref{kop})

-\textit{cross relations}:\medskip
\begin{eqnarray}
\lbrack \widehat{p}_{k},\widehat{x}_{l}]\ &=&\ -i\delta _{kl}\qquad \qquad
\lbrack \widehat{p}_{0},\widehat{x}_{0}]\ = i \medskip  \notag \\
\lbrack \widehat{p}_{k},\widehat{x}_{0}]\ &=&\ -{\frac{i}{\kappa }}\widehat{p%
}_{k}\qquad \qquad \lbrack \widehat{p}_{0},\widehat{x}_{l}]\ =\ 0\medskip
\newline
\notag \\
\lbrack \widehat{m}_{\lambda \rho },\widehat{\lambda }_{\nu }^{\mu }]\ &=&\
i\left( \delta _{\rho }^{\mu }\widehat{\lambda }_{\lambda \nu }-\delta
_{\lambda }^{\mu }\widehat{\lambda }_{\rho \nu }\right) \medskip \newline
,\qquad \lbrack \widehat{p}_{\mu },\widehat{\lambda }_{\rho }^{\lambda }]\
=\ 0  \label{cross} \\
\lbrack \widehat{m}_{\lambda \rho },\widehat{x}^{\mu }]\ &=&\ i\left( \delta
_{\rho }^{\mu }\widehat{x}_{\lambda }-\delta _{\lambda }^{\mu }\widehat{x}%
_{\rho }\right) +{\frac{i}{\kappa }}\left( \delta _{\rho }^{0}\widehat{m}{%
_{\lambda }}^{\mu }-\delta _{\lambda }^{0}\widehat{m}{_{\rho }}^{\mu }\right)
\notag
\end{eqnarray}%
where $\widehat{m}{_{\lambda }}^{\mu }\ =\ g^{\mu \rho }\widehat{m}{%
_{\lambda \rho }},$ $\widehat{m}{^{\mu }}_{\lambda }\ =\ g^{\mu \rho }%
\widehat{m}{_{\rho \lambda }}$.

The generalized covariant $\kappa $-deformed phase space is described by
sets of commutators (\ref{kapp}),(\ref{grup}) and (\ref{cross}). The
coproducts (\ref{kapp2}) and (\ref{kop}) realize the coalgebraic
homomorphism of relations (\ref{kapp}) and (\ref{grup}), but the relations (%
\ref{cross}) will be mapped into the coalgebra only in the bialgebroid
framework (see Sect. 3).

\subsection{Standard covariant $\protect\kappa $-deformed quantum phase space%
}

\bigskip One can obtain the following distinguished subalgebras of general $\kappa $%
-deformed quantum phase space $\mathcal{H}^{(10,10)}$:

\begin{enumerate}
\item By putting consistently in formulae (\ref{grup})-(\ref{cross}) the
value $\widehat{\lambda }{^{\mu }}_{\nu }=\delta {^{\mu }}_{\nu }$,
one obtains the covariant $\kappa $-deformed DSR algebra \cite{24},\cite{25}.
This algebra can be written as the semidirect product
$\HH\rtimes \widetilde{\mathcal{T}_{\kappa }^{4}}$,
with the base generators $\widehat{p}_{\mu },\widehat{m}%
_{\mu \nu },\widehat{x}^{\mu }$, or due to the duality $\mathcal{T}%
^{4}\longleftrightarrow \widetilde{\mathcal{T}_{\kappa }^{4}}$, as
the semidirect product $so(3,1)\rtimes (\mathcal{T}^{4}\oplus \widetilde{%
\mathcal{T}_{\kappa }^{4}})$. The last formula confirms that the $\kappa $%
-deformed Hopf-algebraic Lorentz sector of $\HH$ acts covariantly on the
standard $\kappa $-deformed quantum phase space $\mathcal{H}^{(4,4)}\equiv
\mathcal{H}_{(p,x)}=\HH_{p}\rtimes \widetilde{\HH}_{x}$ defined by relations (%
\ref{7}),(\ref{7a}).

\item If we remove from the general covariant $\kappa $-deformed phase
space\ $\mathcal{H}^{(10,10)}=\HH\rtimes \widetilde{\HH}$ the generators $%
\widehat{m}_{\mu \nu }$, we obtain in a consistent way the algebra with
generators ($\widehat{p}_{\mu },\widehat{x}^{\mu },\widehat{\lambda }_{\mu
\nu }$) which is dual to the considered above $\kappa $-deformed DSR algebra.

\item By removing the Lorentz sector from both Hopf algebras $\HH$ and $%
\widetilde{\HH}$, one obtains the Heisenberg double $\mathcal{H}^{(4,4)}\equiv
\mathcal{H}_{(p,x)}$ with the algebra sector $\mathcal{T}^{4}\oplus
\widetilde{\mathcal{T}_{\kappa }^{4}}$ and the following basic commutators%
\begin{eqnarray}
\lbrack \widehat{x}^{\mu },\widehat{x}^{\nu }]\ &=&\ {\frac{i}{\kappa }}%
(\delta _{0}^{\mu }\widehat{x}^{\nu }-\delta _{0}^{\nu }\widehat{x}^{\mu
})\medskip \newline
\notag \\
\lbrack p_{\mu },p_{\nu }] &=&0\medskip \newline
\label{ph} \\
\lbrack \widehat{p}_{k},\widehat{x}_{l}]\ &=&\ -i\delta _{kl}\qquad \qquad
\lbrack \widehat{p}_{0},\widehat{x}_{0}]\ =i  \notag \\
\lbrack \widehat{p}_{k},\widehat{x}_{0}]\ &=&\ -{\frac{i}{\kappa }}\widehat{p%
}_{k}\qquad \qquad \lbrack \widehat{p}_{0},\widehat{x}_{l}]\ =\ 0.\medskip
\newline
\notag
\end{eqnarray}%
Relations (\ref{ph}) describe the standard $\kappa $-deformed quantum phase
space. For $\kappa \rightarrow \infty $ we get the relativistic quantum
phase space described by the canonical Heisenberg commutation relations. The
relations (\ref{ph}) can not be lifted in homomorphic way to the coalgebra
sector, i.e.\ these relations can not be treated as describing an algebraic
sector of a Hopf algebra\footnote{%
One can add that there were efforts to describe the canonical or deformed
Heisenberg algebra in the framework of Hopf algebras (see e.g.~\cite{26}-%
\cite{28}), but these proposals were in conflict with the basic physical
postulate of standard QM that Planck constant $\hbar $ is an universal
numerical constant, e.g.\ the same for all multiparticle states. Such
impossibility of providing Hopf-algebraic \ framework is valid for all Lie
algebras centrally extended by numerical central charges.}.
\end{enumerate}

\section{Hopf algebroid structure of standard $\protect\kappa $-deformed
quantum phase space}

\subsection{The calculation of target map}

In this section we shall show that using only relations (\ref{1}) one can
determine for $\mathcal{H}^{(4,4)}$ the ideal (\ref{3}). We choose the
following bialgebroid coproducts in $\mathcal{H}^{(4,4)}$, where base
algebra is given by $\widetilde{\mathcal{T}_{\kappa }^{4}}$\ spanned by
generators $\widehat{x}^{\mu }$%
\begin{eqnarray}
\Delta (\widehat{x}^{\mu })\ &=&\ 1\otimes \widehat{x}^{\mu },  \label{bak}
\\
\Delta (\widehat{p}_{k})\ &=&\ \widehat{p}_{k}\otimes e^{-{\frac{\widehat{p}%
_{0}}{\kappa }}}\ +\ I\otimes \widehat{p}_{k},\quad \Delta (\widehat{p}%
_{0})\ =\ \widehat{p}_{0}\otimes I\ +\ I\otimes \widehat{p}_{0}\medskip
\newline
.  \notag
\end{eqnarray}

The coproducts satisfy the commutation relations (\ref{ph})
in $\mathcal{H}^{(4,4)}\otimes \mathcal{H}^{(4,4)}$;
we assume the canonical choice $\alpha (a)=a$. We consider the
relations (\ref{1}) by choosing

\begin{enumerate}
\item $a=\widehat{x}_{0},b=\widehat{x}_{i}\longleftrightarrow \lbrack
\widehat{x}_{0},\beta (\widehat{x}_{i})]=0$ (where $i=1,2,3$)

Choosing $\beta (\widehat{x}_{i})=f(p) \widehat{x}_{i}$ equation
(\ref{ph}) implies%
\begin{equation}
\beta (\widehat{x}_{i})=e^{-{\frac{\widehat{p}_{0}}{\kappa }}}\widehat{x}_{i},
\label{1f}
\end{equation}

\item $a=\widehat{x}_{i},b=\widehat{x}_{0}\longleftrightarrow \lbrack
\widehat{x}_{i},\beta (\widehat{x}_{0})]=0$

Choosing $\beta (\widehat{x}_{0})=\widehat{g}(p)\widehat{x}_{0}+\widehat{g_{i}}(p)\widehat{x}_{i}$ (sum over $i$ understood) we obtain from (\ref{ph})%
\begin{equation}
\beta (\widehat{x}_{0})=\widehat{x}_{0}-\frac{1}{\kappa }\widehat{p}_{i}\widehat{x}_{i}.  \label{f2}
\end{equation}
\end{enumerate}

It is easy to check that for the choice (\ref{1f})-(\ref{f2}) also the
remaining set of eq. (\ref{1}) \ $[\widehat{x}_{0},\beta (\widehat{x}_{0})]=[\widehat{x%
}_{i},\beta (\widehat{x}_{j})]=0$ are valid. Further it can be shown that
the construction of the target map as antihomomorphism
is consistent with the algebraic relations (\ref{4}), i.e.%
\begin{equation}
\lbrack \beta (\widehat{x}_{0}),\beta (\widehat{x}_{i})]=\frac{i}{\kappa }%
\beta (\widehat{x}_{i}),\qquad \lbrack \beta (\widehat{x}_{i}),\beta (%
\widehat{x}_{j})]=0.
\end{equation}

\subsection{The algebraic derivation of coproduct gauge freedom and
bialgebraic equivalence classes}

In this subsection we shall consider arbitrariness of coproducts satisfying
the relations (\ref{ph}) by starting with the formula%
\begin{equation}
\widetilde{\Delta }(\widehat{x}_{\mu })\ =\Delta (\widehat{x}_{\mu })\
+\Lambda _{\mu }(\widehat{x},\widehat{p})=\widehat{x}_{\rho }\otimes \theta
_{\mu }^{\rho }(\widehat{p}),  \label{mcop}
\end{equation}%
where $\Delta (\widehat{x}^{\mu })\ =\ 1\otimes \widehat{x}^{\mu }$ and $%
\theta _{\mu }^{\rho }(\widehat{p})$ is the tensor to be determined. The
relation (\ref{mcop}) describes the homomorphism of -deformed quantum phase
space algebra (\ref{ph}) if the $\Lambda $-tensor operators $\Lambda _{\mu
}\in \mathcal{H}^{(4,4)}\otimes \mathcal{H}^{(4,4)}$ satisfy the relations%
\footnote{%
We denote the algebraic relations described by (\ref{7a}) as $[\widehat{x}%
_{\mu },\widehat{x}_{\nu }]=C_{\mu \nu }^{(\kappa )\ \rho }\widehat{x}_{\rho
}$, where $C_{\mu \nu }^{(\kappa )\ \rho }=\frac{1}{\kappa }(\delta _{\mu
}^{0}\eta _{\nu }^{\rho }-\delta _{\nu }^{0}\eta _{\mu }^{\ \rho })$.}%
\begin{eqnarray}
\lbrack \Delta (\widehat{x}_{[\mu }),\ \Lambda _{\nu ]}]+[\Lambda _{\mu
},\Lambda _{\nu }] &=&C_{\mu \nu }^{(\kappa )\ \rho }\Lambda _{\rho },
\label{r1} \\
\lbrack \Delta (\widehat{p}_{\mu }),\ \Lambda _{\nu }] &=&0.  \label{r2}
\end{eqnarray}

The relations (\ref{r1})-(\ref{r2}) are required if the transformation $%
\Delta (\widehat{x}_{\mu })\longrightarrow \widetilde{\Delta }(\widehat{x}%
_{\mu })$ is todescribe the coproduct gauge. Postulating that $(\widetilde{%
\Delta }(\widehat{x}_{\mu }),\Delta (\widehat{p}_{\mu }))$ satisfies the
quantum phase space algebra relations (\ref{ph}) one derives algebraically
the formulae fixing the tensor $\theta _{\mu }^{\rho }(\widehat{p})$%
\begin{eqnarray}
\widetilde{\Delta }(\widehat{x}_{i})\ &=&\widehat{x}_{i}\otimes e^{\frac{%
\widehat{p}_{0}}{\kappa }},  \label{koa} \\
\widetilde{\Delta }(\widehat{x}_{0})\ &=&\widehat{x}_{0}\otimes 1+\frac{1}{%
\kappa }\widehat{x}_{i}\otimes e^{\frac{\widehat{p}_{0}}{\kappa }}\widehat{p}%
_{i}.  \label{kob}
\end{eqnarray}%
As follows from (\ref{mcop})
one gets\footnote{Similar tensors
$R_{\mu }=\widehat{x}_{\mu }\otimes 1-
\tilde\theta_{\mu }^{\rho}(\widehat{p})\otimes \widehat{x}_{\rho }$,
where $\tilde\theta_{\mu }^\nu$
is the matrix inverse to $\theta_{\lambda }^\rho$ (see~(\ref{mcop})),
have been introduced in \cite{1}
for canonical twisted Heisenberg algebra and considered in \cite{3}-\cite{4}
for $\kappa $-deformed quantum phase space generated by
$\kappa $-deformed Poincar\'e-Hopf algebra
with classical Poincar\'e algebra sector.}%
\begin{eqnarray}
\Lambda _{i} &=&\widehat{x}_{i}\otimes e^{\frac{\widehat{p}_{0}}{\kappa }%
}-1\otimes \widehat{x}_{i},  \label{rr1} \\
\Lambda _{0} &=&\widehat{x}_{0}\otimes 1-1\otimes \widehat{x}_{0}+\frac{1}{%
\kappa }\widehat{x}_{i}\otimes e^{\frac{\widehat{p}_{0}}{\kappa }}\widehat{p}%
_{i}.  \label{rr2}
\end{eqnarray}

One can check subsequently that the relations (\ref{r1})-(\ref{r2}) are
satisfied; in particular in place of eq. (\ref{r1}) we get two equations%
\begin{equation}
\lbrack \Lambda _{\mu },\Lambda _{\nu }]=C_{\mu \nu }^{(\kappa )\ \rho
}\Lambda _{\rho },\qquad \lbrack \Delta (\widehat{x}_{[\mu }),\ \Lambda
_{\nu ]}]=0.  \label{w1}
\end{equation}%
In the limit $\kappa \longrightarrow \infty $ we obtain that $\widetilde{%
\Delta }(\widehat{x}_{\mu })\ =\widehat{x}_{\mu }\otimes 1$.

Further we observe that

\begin{enumerate}
\item In relations (\ref{mcop}) \ one can replace $\Lambda _{\mu
}\longrightarrow \alpha \Lambda _{\mu }$ ($\alpha $-arbitrary constant)
without changing the relations (\ref{r2}) and (\ref{w1}). The resulting
coproducts%
\begin{eqnarray}
\widetilde{\Delta }_{(\alpha )}(\widehat{x}_{i})\ &=&(1-\alpha )(1\otimes
\widehat{x}_{i})+\alpha \widehat{x}_{i}\otimes e^{\frac{\widehat{p}_{0}}{%
\kappa }},  \label{koa1} \\
\widetilde{\Delta }_{(\alpha )}(\widehat{x}_{0})\ &=&(1-\alpha )(1\otimes
\widehat{x}_{0})+\alpha (\widehat{x}_{0}\otimes 1+\frac{1}{\kappa }\widehat{x%
}_{i}\otimes e^{\frac{\widehat{p}_{0}}{\kappa }}\widehat{p}_{i}),
\label{kob1}
\end{eqnarray}

provide parameter-dependent coproduct gauges.

\item Let us replace \ in (\ref{mcop}) $(k\geq 1)$
\begin{equation}
\Lambda _{\mu }\longrightarrow \Lambda _{\mu }^{(k)}\equiv A_{\mu }^{\quad
\nu _{1}\ldots \nu _{k}}\Lambda _{\nu _{1}}\ldots \Lambda _{\nu _{k}}.
\label{chang}
\end{equation}

In such a way we enlarge the class of possible coproduct gauges to any power
of exchange tensor $\Lambda_{\mu }$. Because, using short-hand notation {}%
\begin{eqnarray}
\lbrack \Lambda ^{(k)},\Lambda ^{(l)}] &\subset &\Lambda ^{(k+l-1)},
\label{sh1} \\
\lbrack \widetilde{\Delta }(\widehat{x}_{\mu }),\Lambda ^{(k)}] &=&0,
\label{sh2}
\end{eqnarray}

we see that the homomorphism of coproducts remains valid modulo possible
change of coproduct gauge (\ref{chang}), i.e.\ the algebra (\ref{ph}) is
satisfied by coproducts in equivalence class which is defined by the gauge
freedom described by the basis (\ref{chang}).

\item Finally we insert in (\ref{mcop}) the 2-tensor $\Lambda ^{(k,l,m)}$
depending as well on any power of space-time coproducts and powers of
fourmomenta coproducts ($k\geq 1,l\geq 0,m\geq 0$)%
\begin{equation}
\Lambda _{\mu }\longrightarrow \Lambda _{\mu }^{(k,l,m)}\equiv A_{\mu
}^{\quad \nu _{1}\ldots \nu _{k};\rho _{1}\ldots \rho _{l};\sigma _{1}\ldots
\sigma _{m}}\Lambda _{\nu _{1}}\ldots \Lambda _{\nu _{k}}\Delta (\widehat{x}%
_{\rho _{1}})\ldots \Delta (\widehat{x}_{\rho _{l}})\Delta (\widehat{p}%
_{\sigma _{1}})\ldots \Delta (\widehat{p}_{\sigma _{m}}).  \label{chang22}
\end{equation}

After such substitution and using the properly that ($\Delta (\widehat{x}%
_{\mu }),\Delta (\widehat{p}_{\mu })$) satisfy the algebra\ (\ref{ph}) one
can show that the coproducts ($\widetilde{\Delta }(\widehat{x}_{\mu
}),\Delta (\widehat{p}_{\mu })$) satisfy as well the algebra (\ref{ph}), but
modulo coproduct gauges with the basis (\ref{chang22}).

\item Further, one can introduce analogous coproduct gauge freedom in the
fourmomenta coproducts (\ref{7})%
\begin{equation}
\widetilde{\Delta }(\widehat{p}_{\mu })=\Delta (\widehat{p}_{\mu })+\Lambda
_{\mu }^{(k^{\prime },l^{\prime },m^{\prime })}.
\end{equation}

One can check that the coproducts $\widetilde{\Delta }(\widehat{x}_{\mu }),%
\widetilde{\Delta }(\widehat{p}_{\mu })$ will satisfy the relations (\ref{ph}%
) modulo the gauge freedom spanned by the $2$-tensors given by (\ref{chang22}%
).
\end{enumerate}

The coproduct gauges (\ref{chang22}) define the maximal equivalence class of
coproducts in $\mathcal{H}^{(4,4)}$ inside which the algebra (\ref{ph}) of
coproducts is satisfied. One can say equivalently that the coproduct
gauge-independent description is provided by the equivalence classes of $%
\mathcal{H}^{(4,4)}\otimes \mathcal{H}^{(4,4)}$ which are obtained if we
divide by the ideal with the basis (\ref{chang22}). Analogous ideal using
alternative methods of calculations\footnote{%
Basic tool in \cite{4} providing $\kappa $-deformed coproducts for $\kappa $%
-Minkowski space-time is a deformed Leibniz formula
which describes the action of $\widehat{x}_{\mu }$
on the product $f(\widehat{x})g(\widehat{x})$ of
noncommutative functions on $\kappa $-Minkowski space. In their derivation
besides the algebra (\ref{7a}) of $\kappa $-Minkowski coordinates
some additional input was used
provided by the cross commutators between the $\kappa $-deformed coordinates
and fourmomenta (see also \cite{29}).} has
been recently considered in \cite{4}.

We can show that the coproduct gauge freedom is within the left ideal
$\mathcal{I}_{L}$ given by the relations (\ref{2a}) by noticing
that $\Lambda_\mu(\hat{x},\hat{p}) =
(1\otimes\theta^\rho_\mu)(\alpha(\widehat{x}_\rho)\otimes 1 -
1\otimes\beta(\widehat{x}_\rho))\in\mathcal{I}_L$ (see
(\ref{mcop})).

\subsection{Antipode}

In order to obtain the Hopf algebroid one should supplement the bialgebroid $%
\mathcal{H}$ with an antipode map $\tau:H\longrightarrow H$,
which is a linear antiautomorphism of its total algebra $H$.
It is required \cite{10},\cite{12},\cite{13}
that $\tau$ satisfies the properties which,
in terms of the right bialgebroid, read%
\begin{eqnarray}
\tau\beta  &=&\alpha ,\   \label{f1} \\
m(\tau\otimes\mathrm{id})\widetilde{\Delta } &=&\alpha \epsilon ,  \label{f22} \\
m(\mathrm{id}\otimes \tau)\widetilde{\Delta } &=&\beta \epsilon \tau.  \label{f33}
\end{eqnarray}%
(\ref{f1}) implies that $m(\mathrm{id}\otimes\tau)(\mathcal{I}_L)=0$ hence the
left hand side of~(\ref{f33}) does not depend on coproduct gauge.
In general, however, $m(\tau\otimes\mathrm{id})(\mathcal{I}_L)\neq 0$,
hence the formula (\ref{f22}) can be valid only for the subclass $%
\widetilde{\Delta }_{\gamma }$ of coproducts obtained for specific restrictions of coproduct gauges (see Sect.~3.2). J-H. Lu \cite{10} makes a choice
of linear section $\gamma:H\otimes_A H\to H\otimes H$ such that for
an abstract coproduct $\Delta:H\to H\otimes_A H$ the map $\Delta_\gamma = \gamma\circ\Delta:H\to H\otimes H$ is a specific gauge for which~(\ref{f22}) holds. In \cite{4} a subalgebra $\mathcal{B}$ in $H\otimes H$ is singled out within which all gauge choices satisfy~(\ref{f22}).

For our momentum sector $\mathcal{T}^{4}$,
the antipode $\tau$ is chosen to agree with the Hopf-algebraic antipode:
$\tau(\widehat{p}_0) = S(\widehat{p}_{0})=-\widehat{p}_{0}$ and
$\tau(\widehat{p}_i) =
S(\widehat{p}_{i}) = -e^{{\frac{\widehat{p}_{0}}{\kappa }}}\widehat{p}_{i}$
(see (\ref{kapp3})).
For the coordinate sector, we solve the equations
$\tau(\beta(\widehat{x}_\mu)) = \alpha(\widehat{x}_\mu) = \widehat{x}_\mu$
using that $\tau$ is an antihomomorphism of algebras
and inserting its values on $\widehat{p}_i$. We obtain
\begin{eqnarray}
\tau(\widehat{x}_{i}) &=&e^{-\frac{\widehat{p}_{0}}{\kappa }}\widehat{x}%
_{i}=\tau^{-1}(\widehat{x}_i)=\beta (\widehat{x}_{i}), \\
\tau(\widehat{x}_{0}) &=&\widehat{x}_{0}-\frac{1}{\kappa }\widehat{x}_{i}%
\widehat{p}_{i}=\beta (\widehat{x}_{0})-\frac{3i}{\kappa},
\end{eqnarray}%
\begin{equation}
\tau^{2}(\widehat{p}_{\mu })=\widehat{p}_{\mu },\,\,\,\,\,\,\,\,\,\,\,\,\,
\tau^{2}(\widehat{x}_i)=\widehat{x}_i,\,\,\,\,\,\,\,\,\,\,\,\,\,
\tau^{2}(\widehat{x}_0)=\widehat{x}_0 - \frac{3i}{\kappa}.
\end{equation}%
The counit $\epsilon$ satisfies defining equations
$h_{(1)}\alpha(\epsilon(h_{(2)})) = h = h_{(2)}\beta(\epsilon(h_{(1)}))$
and on generators is given by%
\begin{equation}
\epsilon (\widehat{x}_{\mu })=\widehat{x}_{\mu },\qquad \epsilon (\widehat{p}%
_{\mu })=0,\qquad \epsilon (1)=1.
\end{equation}%
Counit is not an algebra homomorphism but satisfies weaker properties
\cite{12},\cite{13}
\begin{equation}
\epsilon \left(\alpha (\epsilon(h))h'\right) =
\epsilon(h h') = \epsilon \left(\beta (\epsilon(h)) h'\right)\qquad
\epsilon (\alpha (\widehat{x}_{\mu }))=\epsilon
(\beta (\widehat{x}_{\mu }))=\widehat{x}_\mu,
\end{equation}%
and $(f,h)\mapsto \epsilon(\alpha(f)h)$ is a right action, where
$f\in \widetilde{\mathcal{T}_{\kappa}^{4}}$ and
$h,h'\in\mathcal{H}^{(4,4)}=\mathcal{T}^{4}\oplus \widetilde{\mathcal{T}_{\kappa}^{4}}$.

\section{\protect\bigskip Discussion}

The noncommutative Hopf algebras are useful as the tool describing quantum
symmetry algebras and quantum groups \cite{30}-\cite{32}. The notion of Hopf
algebroids at present is well understood as mathematical structure however
its possible physical applications still have to be explored.

An important question in the bialgebroid framework is the physical meaning of
the coproduct freedom, which describes the coproduct gauge.

The momentum sector is described in standard Hopf-algebraic way, with
momenta coproducts $\Delta (p_{\mu })$ describing total 2-particle momentum.
For the coordinates we know how to interpret physically only in very
special case if we deal with nonrelativistic phase space ($%
x_{i}^{(a)},p_{i}^{(a)};$ $i=1,2,3;$ $a=1,2$). For undeformed case ($\kappa
\longrightarrow \infty $) \ the coproduct gauge freedom (\ref{koa1})-(\ref%
{kob1}) looks as follows

\begin{equation}
\Delta (\widehat{x}_{i})=\alpha \widehat{x}_{i}\otimes 1+(1-\alpha )1\otimes
\widehat{x}_{i}.
\end{equation}%
It can be interpreted as characterizing nonrelativistic center-of-mass
coordinate $\widehat{x}_{i}^{(1+2)}$

\begin{equation}
\widehat{x}_{i}^{(1+2)}=\frac{m_{1}}{m_{1}+m_{2}}\widehat{x}_{i}^{(1)}+\frac{%
m_{2}}{m_{1}+m_{2}}\widehat{x}_{i}^{(2)},
\end{equation}%
if we put $\alpha =\frac{m_{1}}{m_{1}+m_{2}}.$ We see therefore that the
coproduct gauge (parameter $\alpha $) is fixed by dynamical parameter of
particles. The values of those parameters is not reflected in the algebraic
relations satisfied by total 2-particle momenta and center-of-mass
coordinates.

Unfortunately, the center-of-mass coordinate for a pair of relativistic
particles (see e.g.~\cite{40},\cite{41}) leads to more complicated
energy-dependent formula for center-of-mass coordinates what does not permit
an analogous interpretation.

In this paper we considered the quantum $\kappa $-deformed phase space
calculated in the Majid-Ruegg basis \cite{7}, with the bicrossproduct
structure of $\kappa $-deformed Poincar\'e algebra successfully applicable for
covariance properties. Such bicrossproduct structure remains valid as well
if we introduce in $\mathcal{T}^{4}$(see (\ref{6a})) an arbitrary
fourmomentum basis%
\begin{equation}
\widehat{p}_{\mu }\longrightarrow \widehat{p}_{\mu }^{\prime }=F_{\mu }(%
\widehat{p}).  \label{pp}
\end{equation}

In particular (see e.g.~\cite{33}-\cite{35}) one can choose the
transformation (\ref{pp}) in a way leading to the classical Poincar\'e basis,
and obtain bicrossproduct structure of $\kappa $-Poincar\'e-Hopf algebra with
classical algebra basis (see e.g.~\cite{39}). The corresponding Heisenberg
double will provide different Hopf algebroid formulae for $\kappa $%
-covariant quantum phase space, with classical action of Lorentz generators
on fourmomenta but complicated fourmomenta coproducts.

One can address the well-known problem in Hopf-algebraic description of
quantum symmetries how to select some priviledged algebra bases of the
bialgebroid. In the case of standard $\kappa $-deformed quantum phase space
one can choose the class of $\kappa $-Poincar\'e algebra bases obtained by
linear choice $F_{\mu }(p)=a_{\mu }^{\nu }p_{\nu }$\ which via Heisenberg
double construction provide class of quantum phase spaces described by $%
\kappa $-deformed centrally extended $8$-dimensional Lie algebras. We add in
particular that if we choose $\kappa $-Poincar\'e algebra with classical
algebra basis, the corresponding quantum $\kappa $-deformed phase space
algebra is not described by such Lie-algebraic formula (see e.g.~\cite{39}).

\bigskip

\textbf{Acknowledgements}

The authors would like to thank A. Borowiec, T. Brzezi\'nski, T. Juri\'c,
D. Kova\v{c}evi\'c, S. Meljanac and P. Kosi\'nski for discussions,
correspondence and valuable remarks.

J.~L. and M.~W. have been supported by
Polish National Science Center projects:
2013/09/B/ST2/02205 and 2014/13/B/ST2/04043.

\end{document}